\documentstyle[12pt]{article}
\begin{document}

\vspace{1.25in}

\centerline{\large\bf Exclusive Scattering at ELFE
\footnote{to be published in the proceedings of the Journees de
Physique Hadronique, Superbesse 1995, ed. Frontieres}}

\vspace{.75in}
\centerline{\large Bernard Pire}
\centerline{\large\it Centre de Physique Th\'eorique}
\centerline{(Unit\'e propre UPR014 du CNRS)}
\centerline{\large\it Ecole Polytechnique, F91128 Palaiseau Cedex, France}
\vspace{1.25in}
\baselineskip=24pt

\centerline{\bf RESUME:}
Il \'etait une fois dans le royaume {\it Standard} un preux chevalier
r\^evant de conqu\'erir une myst\'erieuse province appel\'ee {\it confinement}.
Sa qu\^ete durait depuis plusieurs dizaines d'ann\'ees. Un jour, un
myst\'erieux
{\it elfe} vint lui proposer son aide...C'est le premier chapitre de ses
aventures
que nous ouvrons aujourd'hui, dans ce livre situ\'e aux {\it fronti\`eres} de
cette
 myst\'erieuse province.

\vspace{.75in}

\centerline{\bf ABSTRACT:}
The theoretical framework of hard exclusive reactions is reviewed
with special emphasis on the Elfe project program. Perturbative QCD studies
have
shown that factorization properties allow to separate well-defined non
perturbative
objects which are crucial in the understanding of confinement dynamics from
perturbatively calculable hard processes. The applicability of this
factorization
in a definite energy domain is controlled by some definite statements, as the
dimensional counting rules, the helicity conservation law and the appearance
of color transparency. The few data available indicate that the Elfe parameters
indeed correspond to this well defined physics.

\newpage

\section{\bf Introduction}

\hspace {\parindent}
The theory of hard elastic scattering in Quantum
Chromodynamics (QCD) has evolved considerably over many years of work.
Currently there exists a self-consistent perturbative description, with
a specific factorization method for separating the hard scattering from
non-perturbative wave functions.   A well-known procedure using the
``quark-counting'' diagrams \cite{ref:BrodskyFarrar}{} has been given
 by LePage and Brodsky \cite{ref:pion-ff}{}.
 The value of \(Q^{2}\) at which the scaling
 limit  appears is matter of debate and of detailed
phenomenological analysis.

 The magnetic proton form factor (Fig.1) is the
best known exclusive observable. Scaling obviously settles at \(Q^{2}\) around
10 \( GeV^{2}\).Note that Fig.1 has a linear scale; the slight decrease
at large $Q^2$ is understood as coming from logarithmic corrections which
are well under control (see below). The analysis of the pion form factor is
subject to more
model dependent analysis. It however seems to scale at even lower
 values of \(Q^{2}\).

\begin{figure}
\vspace{4.2in}

\centerline{\small  Fig.1: The proton magnetic form factor}
\end{figure}

In the case of the Deuteron, Brodsky and Chertok
 \cite{ref:BrodskyChertok}{}  propose a very early scaling.
 A recent analysis\cite{ref:FarHul}{}  shows however
that if this is the case then
 the deuteron distribution amplitude is very different from a
 proton-neutron bound state, for instance that it contains, and even
 is dominated by, non nucleonic components. In the case of exclusive
reactions at fixed angle, for instance real or virtual Compton
Scattering or meson photo- and electro-production, the same
arguments predict power behaved cross sections following QCD improved
counting rules. The measure of the \(Q^{2}\) or s-
 dependence of as many as possible hard exclusive observables
 at high energy is thus of
 utmost importance for the understanding of the applicability of
the QCD framework at accessible energies.

A supplementary tool to disantangle the high transfer regime from the
low energy domain is the predicted occurence of color transparency
 when reactions on nuclei are performed. There is still no
decisive experimental evidence of this phenomenon although the
admittedly controversial explanation \cite{ref:RP90}{}
 of pp data at BNL \cite{ref:ASC88}{} and recent data on
diffractive heavy vector meson production at Fermilab
 \cite{ref:Fermilab}{} are certainly indicative.

\section{\bf Factorization : The example of form factors}
\hspace {\parindent}
The spacelike form factor measures the ability of a pion to absorb
a virtual photon (carrying a momentum \(q\) with
\(q^{2}=-Q^{2}<0\)) while remaining intact. It is defined by the
formula:
\begin{equation}
\label{def FF}
<\pi(p')|J^{\mu}|\pi(p)>=e_{\pi}.F(Q^{2}).(p+p')^{\mu},
\end{equation}
where \(e_{\pi}\) is the pion electric charge.

In the hard scattering regime, that is when \(Q\) is very high
with respect to the low energy scales of the theory (the QCD
scale $\Lambda$ and the pion mass), Brodsky and Lepage have
motivated the following three step picture for the process
{}~\cite{ref:pion-ff}{}:
\begin{itemize}
\item
the pion exhibits a valence quark-antiquark {``}soft{''}
(see below) state,
\item
which interacts with the hard photon leading to another soft
state,
\item
which forms the final pion.
\end{itemize}

This leads to the convolution formula:
\begin{equation}
\label{convolution}
F(Q^{2})=\psi_{in}*T_{H}*\psi^{*}_{out}
\end{equation}
and the graphical representation of Fig.2.

{\bf The most important feature of this picture is that it separates
hard from soft dynamics}. The amplitude \(T_{H}\), the
{\em interaction}, reflects the hard transformation due to the
absorption of the photon and is hopefully calculable in
perturbative QCD, because the effective couplings are small in
this regime due to the asymptotic freedom. The amplitude $\psi$,
the {\em wave function}, which depends on low energy dynamics
is outside of the domain of applicability of perturbative QCD
and is, at present, far from being fully understood from the
theory. It is however process independent and contains much
information on confinement dynamics. Factorization proofs
legitimate this picture~\cite{ref:pion-ff}.
\begin{figure}
\vspace{3.5in}

\centerline{\small  Fig.2: The factorization of a hard scattering amplitude  }
\end{figure}

\subsection{Sudakov effects}
\hspace {\parindent}
The need of a careful factorization is due to the infrared
behavior of QCD: technically, large logarithms
(\(\sim\ln(Q/\lambda)\)) appear in the renormalized one-loop
corrections to naive {``}tree-graph{''} ($\lambda$ is some
infrared cut off needed to regularize soft and/or collinear
divergences). As in the renormalization procedure, if
factorization holds, these large corrections should be absorbed,
here in the redefinition of the wave function. The proof of
factorization and its consequences upon the wave functions are
studied in the pattern of the renormalization group. Without
entering into a detailed discussion, let us sketch the procedure
(see~\cite{botste89} for more on this leading
logarithms calculation and also for the renormalization group
treatment).
\begin{figure}
\vspace{3.5in}

\centerline{\small Fig.3: The tree graph for the meson form factor  }
\end{figure}

The first step is to compute the naive hard amplitude, that is
consider the tree graph of Fig.3
, and the three
other graphs related to it by C and T symmetries. One finds,
with notations explained on Fig.3:
\begin{eqnarray}
\label{space-like}
T_{H}=16\pi\alpha_{S}C_{F}
{{xQ^{2}}\over{xQ^{2}+{\bf k}^{2}-i\varepsilon}}\hskip 0.265em
{{1}\over{xyQ^{2}+({\bf k}-{\bf l})^{2}-i\varepsilon}},
\end{eqnarray}
where all quark momentum components are kept. Note
that we have done the usual projection onto the pion S wave
state: \(
\psi_{\pi}(p)\propto {{1}\over{\sqrt{2}}}\gamma^{5}p
\hskip -0.167em \hskip -0.167em \hskip -0.167em /\) and
used the C symmetry of the wave function.
\(C_{F}=4/3\) is the color factor, while \(
\alpha_{S}\) is the QCD effective coupling at the renormalization
point $\mu$.

\begin{figure}
\vspace{3.5in}

\centerline{\small Fig.4: One loop graphs for the meson form factor  }
\end{figure}

Let us now examine one loop corrections to \(T_{H}\) In axial gauge,
it turns out that the relevant graphs to consider are those of Fig.4.:
they directly lead to the wave function correction, in the
{``}double logarithms{''} or Sudakov region (namely:
\( \lambda \ll |{\bf q}| \ll u{{Q}\over{\sqrt{2}}} \ll
x{{Q}\over{\sqrt{2}}} \), $u$ and \({\bf q}\) being respectively
the light-cone fraction and transverse gluon momentum relatively
to the pion):

\begin{eqnarray}
\psi^{(1)}(x,{\bf k})={{C_{F}} \over{2 \pi^{2}}}
\int _{\lambda}^{xQ/ \sqrt{2}}
{{d^{2}{\bf q}} \over{{\bf q}^{2}}} \alpha_{S}({\bf q}^{2})
\int _{|{\bf q}| \sqrt{2}/Q}^{x} {{du}\over{u}}
\{\psi^{(0)}(x-u,{\bf k}+{\bf q})
-\psi^{(0)}(x,{\bf k})\} \nonumber
\\
+{{C_{F}} \over{2 \pi^{2}}}
\int _{\lambda}^{\overline{x} Q/ \sqrt{2}}
{{d^{2}{\bf q}} \over{{\bf q}^{2}}} \alpha_{S}({\bf q}^{2})
\int _{|{\bf q}|\sqrt{2}/Q}^{\overline{x}}{{du}\over{u}}
\{ \psi^{(0)}(x+u,{\bf k}+{\bf q})
-\psi^{(0)}(x,{\bf k})\} ,
\label{one-loop}
\end{eqnarray}
where \( \overline{x} =1-x\) and the first term in the
difference comes from vertex-like corrections and the second
one
from self energy ones; in the infrared region some partial
cancellations occur between these corrections, but the
cancellation is not complete.

To pursue this analysis, it is convenient to define the Fourier
transform in the transverse plane:
\begin{equation}
\hat{\psi}{(x,{\bf b})}=\int d^{2}{\bf k}
e^{i{\bf kb}} \psi{(x,{\bf k})},
\end{equation}
and to separate transverse and longitudinal variations of the wave
function. One finds, omitting for the moment the second term in
Eq.~(\ref{one-loop}):
\begin{eqnarray}
\hat{\psi}^{(1)}(x,{\bf b})
={{C_{F}}\over{2\pi^{2}}}
\left(\int {{d^{2}{\bf q}}\over{{\bf q}^{2}}}
\alpha_{S}{({\bf q}^{2})}
(e^{-i{\bf qb}}-1)\int {{du}\over{u}}\right)
\hat{\psi}^{(0)}(x,{\bf b}) \nonumber
\\
+{{C_{F}}\over{2\pi^{2}}}
\int {{d^{2}{\bf q}}\over{{\bf q}^{2}}}\alpha_{S}({\bf q}^{2})
e^{-i{\bf qb}}
\int {{du}\over{u}}\left(\hat{\psi}^{(0)}(x-u,{\bf b})
-\hat{\psi}^{(0)}(x,{\bf b})\right).
\label{one-loop-b}
\end{eqnarray}

This equation contains the typical corrections one has to
consider in a hard process when dealing with either a big
(\(\gg 1/Q\)) or a small ( $< 1/Q $) neutral object.

The transverse behavior at large distance is driven by the first
term of the previous equation, thanks to the vanishing of the
summation with the oscillating components. This occurs when
\(b={|{\bf b}|}\) is greater than a few times the inverse
of the upper bound of the corresponding integral:
\(xQ/\sqrt{2} \). As a consequence, in the remaining expression,
the infrared cut-off $\lambda$ can be replaced by the natural one
\(1/b\), above which the vextex and self energy corrections
do not compensate one another. Thus we get~\cite{li-ste92}:
\begin{equation}
\hat{\psi}^{(1)}=-s(x,Q,b)\hskip 0.265em \hat{\psi}^{(0)},
\hskip 0.265em s={{C_{F}}\over{2\beta}}
\ln \hskip 0.265em {{xQ}\over{\sqrt{2}}}
\left( \ln {{\ln \hskip 0.212em xQ/\sqrt{2}}
\over{\ln \hskip 0.212em 1/b}}
-1+{{\ln \hskip 0.212em 1/b}\over{\ln \hskip 0.212em
xQ/\sqrt{2}}}\right),
\label{one-loop-large-b}
\end{equation}

\noindent
with $\beta=(11-{{2n_{f}}\over{3}})/4\), \(n_{f}$ being
the number of quark flavors. Here and in the following, it is
understood that the energies and inverse separations
are in units of the natural QCD scale $\Lambda _{QCD}$.
After the ressummation of the ladder structure to all orders, the
above Sudakov factor exponentiates.
Taking into account the term
obtained with the substitution
$x \rightarrow \overline{x} \equiv 1-x$, we get:
\begin{equation}
\hat{\psi} (x,b,Q)=e^{-s(x,b,Q)-s(\bar{x},b,Q)}
\hat{\psi} ^{(0)}(x,b,Q).
\end{equation}

Thus we get a strong suppression of the effective wave function as
\(b\rightarrow 1/\Lambda\), whatever the fraction $x$ is, provided
that $Q$ is reasonably large.
{\bf The remaining object $\hat{\psi}^{(0)}$  is a soft component to
start with}. It is soft in the sense that it does not include
loop-corrections harder than $1/b$. One may modelize it by including
some $b$ behavior or simply relate it to the distribution
amplitude~\cite{botste89} setting:
\begin{equation}
\hat{\psi}^{(0)}(x,b) \approx
\int _{0}^{1/b}\psi{(x,{\bf k})}d{\bf k}=\varphi(x;1/b).
\end{equation}

\subsection{Leading log analysis}
\hspace {\parindent}
The first term in Eq.6 is negligible when
the oscillating term remains close to $1$ in the range of
integration. This happens for $b$ a few times less than
\(\max ^{-1}(xQ,\overline{x}Q)\). In this case, soft divergences
cancel one another and one finds:
\begin{equation}
\label{10}
\hat{\psi}^{(1)}(x)
=\xi {{C_{F}}\over{2}}
\int _{0}^{1}dx' \left\{ {{\hat{\psi}^{(0)}(x')-
\hat{\psi}^{(0)}(x)}\over{x-x'}}
\theta (x-x')
+{{\hat{\psi}^{{(0)}}(x')-\hat{\psi}^{(0)}(x)}\over{x'-x}}
\theta (x'-x) \right\} ,
\end{equation}
\noindent
with the notation:
\begin{equation}
\label{11}
\xi={{1}\over{\pi^{2}}}
\int _{\lambda}^{Q}{{d^{2}{\bf q}}\over{{\bf q}^{2}}}
\alpha _{S}({\bf q}^{2})
\sim {{1}\over{\beta}} \ln \left({{\ln Q}\over{\ln \lambda}} \right).
\end{equation}

We displayed this equation in a slightly different form than in the
large $b$ case to explicitly show that equation \ref{one-loop}, in
the limit of small $b$, is related to the distribution evolution
proposed in~\cite{ref:pion-ff}.
Let us shortly review how this comes\cite{ref:GP}. The earlier simpler
factorization formula  for exclusive
processes~\cite{ref:pion-ff} is easily derived from the
previous treatment if one assumes that neither the wave function
nor the hard amplitude give important contribution to the
form~factor when the transverse momenta are big. Neglecting all
transverse momenta in $T_{H}$ leads therefore to consider the
$k_{T}$-integrated quantity:

\begin{equation}
\varphi(x)=\int d^{2}{\bf k} \psi (x,{\bf k})
\end{equation}

This distribution amplitude related to the wave function at $b=0$
has a dependence in $Q$ associated with the remaining
collinear  divergences
in Eq.6. Indeed, the exponentiated form of
this convolution equation, once it is written for the distribution
$\varphi$ and generalized to other regions than the Sudakov one,
leads to the celebrated expansion of
\(\varphi (x,Q)/x\bar{x}\) in a linear combination of running
logarithms together with Gegenbauer polynomials.

Before turning to this point, let us examine a useful constraint on
 the distribution amplitude. In fact we know something about the wave function,
and this is because
the pion electroweak decay is a {\it very} short distance processus, which
 requires the pion to be in its valence state. The pion decay is described
by the matrix element:

\begin{equation}
\label{13}
\langle0|\bar{q}_d(0)\gamma^{\mu}(1-\gamma^5)q_u(0)|\pi^+(p)\rangle=
f_{\pi}p^{\mu},
\end{equation}
\noindent
where $f_{\pi}$ has been measured as 133 MeV.

The amplitude at zero distance may be written as:
\begin{equation}
\label{14}
\langle0|T\left(q_{u\alpha i}(0)\bar{q}_{d\beta j}(0)\right)|\pi^+(p)\rangle
=\int_0^1dx{Q\over2\sqrt{2}\pi}\int {dk^-d{\bf k}_{\bot}\over (2\pi)^3}X(k),
\end{equation}
\noindent
which must be projected onto the tensor
$\gamma^{\mu}(1-\gamma^5)|_{\beta\alpha}\delta_{ji}$, to yield:

\begin{equation}
\label{15}
-\langle0|T\left(\bar{q}_{d i}(0)\gamma^{\mu}(1-\gamma^5)
q_{u i}(0)\right)|\pi^+(p)\rangle
=\int_0^1\varphi(x)
Tr\left({\gamma^5p\!\!\!/\over4}\gamma^{\mu}(1-\gamma^5)\right)
{\delta_{ij}\over3}\delta_{ji}.
\end{equation}
\noindent
 We get:

\begin{equation}
\int_0^1dx\varphi(x)p^{\mu}=f_{\pi}p^{\mu},
\end{equation}
\noindent
so that the normalisation of the  distribution amplitude is fixed.

Let us now see how leading logarithms are resummed into the distribution
amplitude, in a way which is much reminiscent of the now standard
 Altarelli-Parisi
evolution equations for structure functions in deep inelastic reactions.
In axial gauge, one may isolate big logarithmic factors in ladder-type
 diagrams, leading to an expansion:
\begin{eqnarray}
\varphi_{\rm LL}(x,Q)&=&
\varphi_0(x)
+\kappa\int_0^1du\,V_{q\bar{q}\rightarrow q\bar{q}}(u,x)\,\varphi_0(u)
\nonumber\\
&+&{\kappa^2\over2!}\int_0^1duV_{q\bar{q}\rightarrow q\bar{q}}(u,x)
\int_0^1du'V_{q\bar{q}\rightarrow q\bar{q}}(u',u)\,\varphi_0(u')+\ldots
\end{eqnarray}
\noindent
 where
\[
\kappa={2\over \beta}\ln{\alpha_S(\mu^2)\over\alpha_S(Q^2)},
\]
\noindent
and $V_{q\bar{q}\rightarrow q\bar{q}}$ is the following kernel
\begin{equation}
V_{q\bar{q}\rightarrow q\bar{q}}(u,x)={2\over3}\left\{
{\bar{u}\over\bar{x}}\left(1+{1\over u-x}\right)_+\theta(u-x)+
{u\over x}\left(1+{1\over x-u}\right)_+\theta(x-u)\right\},
\end{equation}
\noindent
where the $()_+$  distribution comes from the colour neutrality of the
meson.

The equation on  $\varphi$ may be rewritten as
\begin{equation}
\left({\partial\varphi\over\partial\kappa}\right)_x=
\int_0^1duV(u,x)\,\varphi(x,Q),
\end{equation}
\noindent
the solution of which has been known for a long time as:
\begin{equation}
\varphi(x,Q)=x(1-x)\sum_n\phi_n(Q)C_n^{(3/2)}(2x-1);
\end{equation}
\noindent
where the $C_n^{(m)}$ are Gegenbauer polynomials which verify:
\begin{equation}
\int_0^1du~u(1-u)\,V(u,x)\,
C_n^{(3/2)}(2u-1)=A_nx(1-x)C_n^{(3/2)}(2x-1),
\end{equation}
\noindent
with eigenvalues $A_n$. One thus gets
\begin{equation}
\phi_n(Q)=\phi(\mu)e^{A_n\kappa}=\phi(\mu)\left(
{\alpha_S(\mu^2)\over\alpha_S(Q^2)}\right)^{2A_n/\beta},
\end{equation}
\noindent
where exponents monotonouly decrease, beginning with:
${2A_0\over\beta}=0,\ {2A_2\over\beta}=-0,62,\ \ldots$

Using the previously derived normalization:
\begin{equation}
\int_0^1dx\varphi(x,Q)=\phi_0(Q)\int_0^1dxx(1-x)={\phi_0\over6}=f_{\pi},
\end{equation}
\noindent
the expansion may be written as:
\begin{equation}
\varphi(x,Q)=6f_{\pi}x(1-x)+\Phi_2(\ln Q^2)^{-0.62}x(1-x)(2x-1)^2+\ldots
\end{equation}

We thus know the asymptotic pion distribution amplitude:
\begin{equation}
\varphi(x,Q)\sim_{Q\rightarrow\infty}6f_{\pi}x(1-x).
\end{equation}
We however do not know the realistic distribution amplitude at accessible
energies, since the constants $\Phi_2,\ldots\Phi_n$ cannot be derived from
perturbative QCD.
This is where the QCD sum rule approach enters \cite{ref:deR}{} to
 yield values for moments of the distribution

\begin{equation}
\int_0^1dx(2x-1)^{2n}\varphi(x,\mu),\ \ldots
\end{equation}
and then reconstruct the constants $\Phi_2,\ldots\Phi_n$ \cite{ref:CZ}{}.

\section{\bf Exclusive reactions at ELFE }

\hspace {\parindent}
Measuring the pion form factor, and more generally the dominant
form factors for hadrons ($ F_K $,$ G_N$, ...) just allows to get
one constraint on  the distribution amplitude of these hadrons. To
go further, one needs to investigate scattering amplitudes at
large angle which map out the $ x_i$ dependence of the wave functions.
Elastic real and virtual Compton scattering ($e+p \rightarrow e+p+\gamma$)
at high momentum transfer meet this goal since they only depend on
the valence proton distribution amplitude (eventually Sudakov improved)
analysed by a calculable hard amplitude. Different meson photo- and
electro-production processes depend on more hadronic wave functions and
are as thus probes of various ways quarks are confined in hadrons.

Subdominant form factors (such as $F_2$,...) and helicity violating processes
(such as non-diagonal helicity matrix elements of produced vector mesons)
are currently not understood with the same degree of rigor. Some progress has
 however been made  \cite{ref:gpr}{} in a related process where it has been
proposed that these observables measure a non-zero orbital angular momentum
 part of the valence wave-function.

\subsection{\bf Compton Scattering }

\hspace {\parindent}
 To lowest order in
the fine structure constant $\alpha \sim 1/137$, Virtual Compton
 Scattering (VCS) is
described by the coherent sum of the amplitudes shown in Fig.5, namely
 a Bethe Heitler process (Fig.5b) where the final photon is radiated off
 the electron and a genuine VCS process (Fig.5a).
\begin{figure}
\vspace{3.5in}

\centerline{\small Fig.5: Amplitudes for virtual Compton scattering  }
\end{figure}
Since the BH amplitude is calculable, at least  once the
 elastic form factors $G_{Mp}(-t)$ and $G_{Ep}(-t)$ are known,
 its interference with VCS
is a new source of information, not present in either real
compton scattering or in electroproduction experiments (e.g.
$(e,e'p)\pi^0$ ). The VCS amplitude depends upon three independent
 invariants, the usual choices being $ Q^2,  s, t $
 or $ Q^2, s, \theta_{CM} $.
\begin{figure}
\vspace{3.5in}

\centerline{\small Fig.6: Real Compton Scattering at large angle  }
\end{figure}

Fig.6  displays the world's supply of high energy
 real Compton data on the proton,
for $-t>1GeV^2$.  Although there are many experiments at high energy
 and low $t$, there is only the experiment of
 Shupe {\it et al.,\/} at large $t$.\cite{ref:Shupe},
The data are plotted as $s^6 d\sigma/dt$ {\it vs.\/} $\cos\theta_{CM}$
to illustrate the approach to the asymptotic scaling law.
The most stringent test of the scaling law is obtained at $\theta_{CM} =
 90^\circ$.
Fitting the data to a $s^{-\alpha}$ power law results in
 $\alpha= 7.0\pm0.4$:
a $ 2.5 \sigma$ deviation from the $\alpha=6.0$ prediction.

The perturbative QCD
calculations of high energy real~\cite{ref:FarrarMaina,ref:Kronfeld}{}
and virtual compton scattering~\cite{ref:FarrarZhang}{}
on the proton keep only the
 lowest order Feynman diagrams (there are 336
non-vanishing topologically distinct ways to couple two photons to
three quarks with the exchange of two gluons and 42 diagrams with
the three gluon vertex whose color factor however vanishes).
Each quark entering or leaving the hard scattering carries a fraction $x$ of
the momentum of its parent proton and
components of momentum along the three other directions. As long as
$x$-fractions are fixed and non-xero , and at large enough momentum
transfer, it is legitimate to further neglect, in the hard amplitude,
components of each quark internal momenta which do not lie along its
parent proton , one thus gets
\begin{equation}
A=\phi_{(uud)}\otimes T_H(\{x\},\{y\})\otimes \phi'_{(uud)}(1+O(M^2/t)),
\end{equation}

\subsection{\bf A strategy for data analysis }
\hspace {\parindent}

A first way to analyse data is to compare experimental points
to a calculation with a given distribution amplitude that any
prejudiced theorist convinced you to choose. This is the way that
the pQCD calculations by Kronfeld and Ni\v zi\'c \cite{ref:Kronfeld}{}
 of real Compton scattering are shown in Fig.6 for various choices of
 the distribution amplitude. One observes the good sensitivity of
the Compton cross section to the non perturbative quantity $ \phi_{(uud)}$
we are primarily intereted in.

A more model-independent way is to try to sort out the wave function
directly from the data. Let us outline a possible strategy on the example
 of real Compton scattering. We first write the proton valence
 wave-function as the series derived from the leading logarithmic
analysis, similar to Eq.24 for the pion case, that is in terms of
 Appel polynomials \cite{ref:CZ}{}, as

\begin{equation}
\phi(x_i,Q)=120 x_1 x_2 x_3 \left\{
1 + {{21}\over{2}}
\left({\alpha_S(Q^2)\over\alpha_S(Q_0^2)}\right)^{\lambda_1} A_1
 P_1(x_i)+ {{7}\over{2}} \left({\alpha_S(Q^2)
\over\alpha_S(Q_0^2)}\right)^{\lambda_2} A_2 P_2(x_i)
+...\right\} ,
\end{equation}
\noindent

where the slow $Q^2$ evolution entirely comes from renormalization group
factors $ \alpha_S(Q^2)^\lambda$, $ \lambda_i$  being calculated increasing
 numbers:
\[
\lambda_1 = {{20}\over{9\beta}}   ~~~~,~~~~\lambda_2 = {{24}\over {9\beta}} ,
\]

\noindent
$P_i(x_j) $ are tabulated Apple polynomials
\[
P_1(x_i)=x_1-x_3 ~~~,~~~ P_2(x_i)=1-3x_2 ,...
 \]
and $A_i$ are unknown constants which measure the projection of the
wave-function on the Apple polynomials:
\begin{equation}
A_i=\int _{0}^{1}dx_1 dx_2 dx_3 ~\delta(x_1+x_2+x_3-1)~\phi(x_i)~P_i(x_i)
\end{equation}

\noindent
We then write the Compton differential cross-section as a sum of terms
\[
A_i T_H^{ij}(\theta) A_j
\]
\noindent
where $T_H^{ij}$ are integrals of the hard amplitude at some given
scattering angle $\theta$ multiplied by the two Apple polynomials
$A_i(x)$ and $A_j(y)$ over light-cone variables x and y, which have
a rather awful analytical expression but can easily be electronically
 stored.

Sorting out the valence wave function of the proton from the data
amounts then to determine through a maximum of likelyhood method the
parameters $A_i$ restricting to something like ten terms in the expansion
of Eq.28. A direct test of the validity of the approach is then to explore
both real and virtual Compton scattering data which should be understood
with the same series of  $A_i$'s.

\subsection{\bf Other processes }
\hspace {\parindent}

Photo- and electro-production of mesons at large angle will enable to
probe $\pi$ and $\rho$ distribution amplitudes in much the same way. The
production of $ K \Lambda $ final states will enable to enter strange
 quark production, thus selecting few diagrams in the hard process. Not
much theoretical analysis of these possibilities has however been worked
out except under the simplifying assumptions of the diquark model
\cite{ref:diquark}{}.

\section{\bf Color Transparency }
\subsection{\bf The idea }
\hspace {\parindent}
The concept of Color  transparency~\cite{ref:MU82,FR92}{}
has recently attracted much attention.
This phenomenon illustrates the power of exclusive reactions to  isolate
simple elementary quark configurations. The experimental technique  to
probe these configurations is the following:

 For  a hard  exclusive reaction,  say electron  scattering from  a
proton, the  scattering amplitude  at large  momentum transfer  $Q^2$ is
suppressed  by  powers  of  $Q^2$  if  the proton contains more than the
minimal number  of constituents.   This  is derived  from the  QCD based
quark  counting   rules,  which   result  from   the  factorization   of
wave-function-like  distribution  amplitudes.    Thus protons containing
only  valence  quarks  participate  in  the  scattering.  Moreover, each
quark,  connected  to  another  one  by  a  hard gluon exchange carrying
momentum of order $Q$, should be found within a distance of order $1/Q$.
Thus , at  large $Q^2$ one  selects a very  special quark configuration:
all connected  quarks are  close together,  forming a  small size  color
neutral  configuration  sometimes  referred  to  as a {\em mini hadron}.
This mini hadron  is not a  stationary state and  evolves to build  up a
normal hadron.

Such  a color  singlet system  cannot emit  or absorb  soft gluons
which carry energy or momentum smaller than $Q$.  This is because  gluon
radiation --- like photon radiation in QED --- is a coherent process and
there is thus destructive interference between gluon emission amplitudes
by quarks  with ''opposite''  color.   Even without  knowing exactly how
exchanges  of   soft  gluons   and  other   constituents  create  strong
interactions, we  know that  these interactions  must be  turned off for
small color singlet objects.

An exclusive hard reaction will thus probe the structure of a {\em  mini
hadron}, i.e. the short distance part of a minimal Fock state  component
in the hadron  wave function.   This is of  primordial interest for  the
understanding  of  the  difficult  physics  of  confinement.    First,
selecting the simplest Fock state amounts to the study of the  confining
forces in  a colorless  object in  the ''quenched  approximation'' where
quark-antiquark pair creation from  the vacuum is forbidden.   Secondly,
letting the mini-state evolve during its travel through different nuclei
of various  sizes allows  an indirect  but unique  way to  test how  the
squeezed mini-state  goes back  to its  full size  and complexity,  {\em
i.e.} how  quarks inside  the proton  rearrange themselves  spatially to
''reconstruct'' a normal size hadron.   In this respect the  observation
of baryonic resonance  production as well  as detailed spin  studies are
mandatory.

To the extent that the electromagnetic form factors are understood as  a
function of $Q^2$, $eA \rightarrow e' (A-1) p$ experiments will measure
the color screening properties of QCD.   The quantity to be measured  is
the transparency ratio $T_r$ which is defined as:
\begin{equation}
T_r = \frac{\sigma_{Nucleus}}{Z \sigma_{Nucleon}}
\end{equation}

At asymptotically large values  of $Q^2$, dimensional estimates  suggest
that $T_r$ scales as a function of $A^{\frac{1}{3}}/Q^2$~\cite{ref:PR}{}.
  The  approach
to the scaling behavior as well as  the value of $T_r$ as a function  of
the  scaling  variable  determine  the  evolution  from  the   pointlike
configuration to the  complete hadron.   This highly interesting  effect
can be measured in an $e , e' p$ reaction that  provides
the best chance for a {\it quantitative} interpretation.
\subsection{\bf Available data }
\hspace {\parindent}
Experimental data on color transparency are very scarce but worth
considering in detail. The first piece of evidence for something like
color transparency came from the Brookhaven experiment on pp elastic
scattering at 90¡ CM in a nuclear medium ~\cite{ref:ASC88}{}. These data
lead to a lively debate with no unanumous conclusion. The problem is
that hadron hadron elastic scattering is not an as-well clear cut case
of short distance process as the electromagnetic form factors discussed
above. There are indeed infrared sensitive processes (the so-called
independent scattering mechanism) which allow not so small protons to
scatter elastically. The phenomenon of colour transparency is thus
replaced by a {\it nuclear filtering} process: elastic scattering in
a nucleus filters away the big component of the nucleon wave function
and thus its contribution to the cross-section. Since the presence of
these  two competing processes had been analysed \cite{ref:PR82}{} as
responsible of the oscillating pattern seen in the scaled cross-section
$s^{10}d\sigma/dt$ , an oscillating color transparency ratio emerges (see
Fig.7)
\begin{figure}
\vspace{3.5in}

\centerline{\small Fig.7: Oscillating scaled cross-section $R_1$ (a)}
 \centerline{\small and
 transparency ratio (b) for pp elastic scattering at $90$
degrees\cite{ref:RP90}.}
\end{figure}

The SLAC NE18\cite{ref:SLAC}{} experiment recently
 measured the color transparency ratio up to
$Q^2 = 7 GeV^2$ , without any observable increase. This casts doubts on the
most
optimistic views on very early dominance of point like configurations and
emphasizes the importance of a sufficient boost to prevent small states to
dress-up to quickly, then losing their ability to escape freely the nucleus.

The diffractive electroproduction of heavy vector mesons at
 Fermilab~\cite{ref:Fermilab}{} recently showed
an interesting increase of the transparency ratio for data at $Q^2 = 7 GeV^2$ .
In this case the boost is high since the lepton energy is around
$E \simeq 200 GeV$ but the problem is to disentangle diffractive
from inelastic events.

\section{\bf Conclusion }
\hspace {\parindent}
Exclusive Scattering is certainly a central issue of the ELFE program and
the beautiful theoretical progresses we have witnessed during these last
few years eagerly await experimental data to be confronted with. We at last
have a chance to pierce the secret of how quarks manage to build up
hadrons by a thorough study of rare but illuminating processes. The new
techniques now available for accelerating intense electron beams with
a very high duty factor are quite fascinating. They will allow us to open a
 new chapter in the study of this challenging object we are all made of,
namely the {\it proton}.

\vspace{.5in}
\noindent {\bf Acknowledgments}:
I thank Bernard Frois, Thierry Gousset, Pierre Guichon, Al Mueller and John
 P Ralston for many discussions.
Centre de Physique Th\'eorique is unit\'e propre du CNRS.

\end{document}